# Automated generation of epilepsy surgery resection masks; The RAMPS pipeline

Callum Simpson[1], Gerard Hall[1], John S. Duncan[3], Yujiang Wang[1,2,3*], Peter N. Taylor[1,2,3*],

1. CNNP Lab (www.cnnp-lab.com), Interdisciplinary Computing and Complex BioSystems Group, School of Computing, Newcastle University, Newcastle upon Tyne, United Kingdom
2. Faculty of Medical Sciences, Newcastle University, Newcastle upon Tyne, United Kingdom
3. UCL Queen Square Institute of Neurology, Queen Square, London, United Kingdom

Corresponding author *peter.taylor@newcastle.ac.uk

## Abstract

MRI-based delineation of brain tissue removed by epilepsy surgery can be challenging due to post-operative brain shift. In consequence, most studies use manual approaches which are prohibitively time-consuming for large sample sizes, require expertise, and can be prone to errors.

We propose RAMPS (Resections And Masks in Preoperative Space), an automated pipeline to generate a 3D resection mask of pre-operative tissue. Our pipeline leverages existing software including FreeSurfer, SynthStrip, Sythnseg and ANTS to generate a mask in the same space as the patient's pre-operative T1 weighted MRI. We compare our automated masks against manually drawn masks and two other existing pipelines (Epic-CHOP and ResectVol).

Comparing to manual masks (N=87), RAMPS achieved a median(IQR) dice similarity of 0.86(0.078) in temporal lobe resections, and 0.72(0.32) in extratemporal resections. In comparison to other pipelines, RAMPS had higher dice similarities (N=62) (RAMPS:0.86, Epic-CHOP: 0.72, ResectVol: 0.72).

We release a user-friendly, easy to use pipeline, RAMPS, open source for accurate delineation of resected tissue.

# 1 Introduction

Surgical removal of the epileptic foci as a treatment for drug resistant epilepsy leads to seizure freedom in only 50% of cases (De Tisi et al., 2011). This fact sparked a wave of retrospective studies focused on re-examining the preoperative tissue under new hypotheses (Ahmed et al., 2015; Galovic et al., 2019; Hall et al., 2023; Horsley et al., 2023; Keller et al., 2015; Sainburg and Morgan, 2024; Taylor et al., 2018), all aiming to uncover novel biomarkers to predict surgical outcome. In all such studies, accurate delineation of the resected tissue is crucial to capture precise information.

A common approach to delineate the resection is manually drawing by hand a resection mask, a 3D binary matrix labeling individual T1-weighted voxels as a part of the resection cavity (Taylor et al., 2018). Manual delineation is time-consuming, requiring a skilled individual with advanced neuroanatomical and surgical knowledge (Arnold et al., 2022; Gau et al., 2020). Even with such knowledge, the manually generated mask varies between raters (Pérez-García et al., 2020).

These issues highlight the need for an automated technique capable of delineating the resection cavity to 1) reduce time investment 2) reduce domain knowledge needed, increasing the accessibility to resection mask based analysis 3) enable comparison of results against additional resection classifications 4) facilitate easier replication of previous studies on new datasets.

Several tools to delineate the resection cavity exist in three categories; deep/machine learning (Arnold et al., 2022; Casseb et al., 2024; Pérez-García et al., 2021), semi-automated (Billardello et al., 2022; Wilke et al., 2011) and fully automated statistical (Cahill et al., 2019; Casseb et al., 2021). Many of these resources generate the mask in the post-operative space by filling in the resection cavity (Figure 1 C). These approaches yield good results when compared to other manual masks also drawn in the post-operative space. However these approaches are prone to errors caused by post-operative sagging, swelling, or brain shift into the resection cavity. Visual comparison of the post-operative mask to a mask delineating the pre-operative tissue resection (Figure 1 D) highlights differences in volume and extent (Figure 1 E). This problem can be further exacerbated if the registration between the two images is poor, causing additional warping to the resection mask. The creation of resection masks in pre-operative, rather than post-operative space is therefore of paramount importance.

Here we present the RAMPS pipeline (Resections And Masks in Pre-operative Space). The pipeline which generates masks of the subsequent resection in pre-operative space. Using a three-step pipeline building on previously available tools we automated the creation of pre-operative resection masks, highlighting the pre-operative tissue shown to be resected in the post-operative image. This is achieved using a patient's pre-operative and post-operative T1w image, information regarding resected lobe and hemisphere information. RAMPS is open access, and all code is freely available to download (https://github.com/cnnp-lab/RAMPS).

# 2 Materials and methods

## 2.1 Data used

In this study we used pre- and post-operative T1-weighted MRI brain scans from 87 individuals who had epilepsy surgery. In each case a manual resection mask was drawn on the pre-operative scan, highlighting the tissue shown to be resected in the post operative scan. To validate the performance under diverse conditions, there was no exclusion criteria when selecting patients. Scans include Temporal Lobe Epilepsy TLE (N=70) and Extra Temporal Lobe Epilepsy ETLE cases (N=17), collected from different centres and 3D T1-weighted images were acquired across different protocols. The majority of the scans were acquired at the Chalfont Centre, with manual masks delineated for previous studies (Hall et al., 2023; Taylor et al., 2018), with additional scans being collected from the University Hospital Iowa (N=2 used by (Kocsis et al., 2023)), Pennsylvania, and Mayo Clinic.

## 2.2 Manual segmentation

All patients had a manually drawn mask highlighting the pre-operative tissue that was subsequently resected. This was achieved by visually comparing the post-operative scans against the pre-operative to delineate the resected tissue, then slice by slice highlighting the pre-operative voxels. These masks were created by 3 individuals trained in this process. For this study, all manual masks were filtered by a binary segmentation of the pre-operative brain tissue to remove erroneous voxels outside the brain (e.g. CSF, bone). If not already in $1mm^3$ resolution, manual masks were converted to match this resolution.

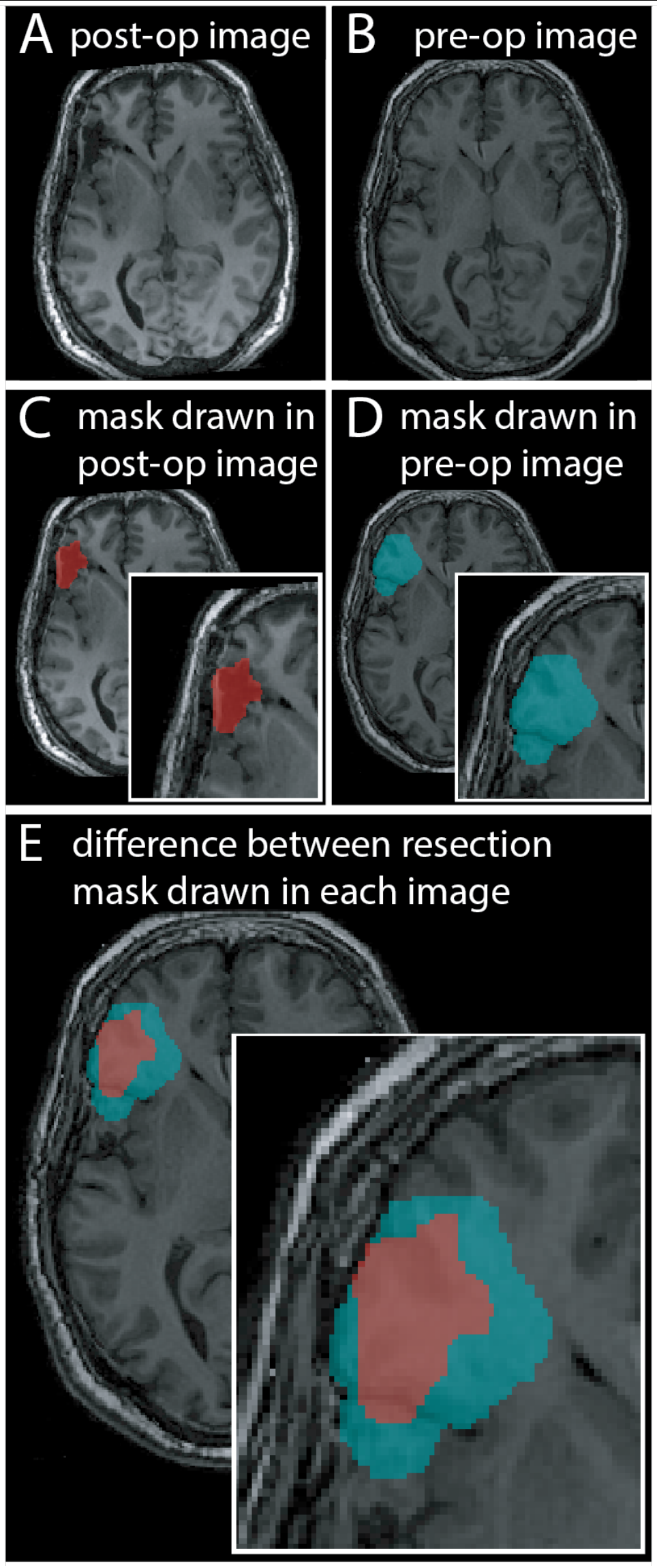

***Figure 1: The need to generate resection masks in preoperative space*** *Illustrated is the axial aligned A) post-operative and B) the pre-operative T1w images for a frontal lobe resection. Following surgery, sagging is noted within the resected cavity. C) shows a manually drawn mask following the resection cavity seen in the post-operative image. D) shows a manually drawn resection mask that indicates the pre-operative tissue that is shown to be resected in the post-operative image. E) Highlights the difference between these two resection cavity interpretations.*

## 2.3 RAMPS pipeline

The RAMPS pipeline comprises 3 core sections: data preparation, registration and mask creation (Figure 2). RAMPS runs in Python and uses modules such as nibabel (Brett et al., 2024) and Ants (Tustison et al., 2021) to aid mask creation.

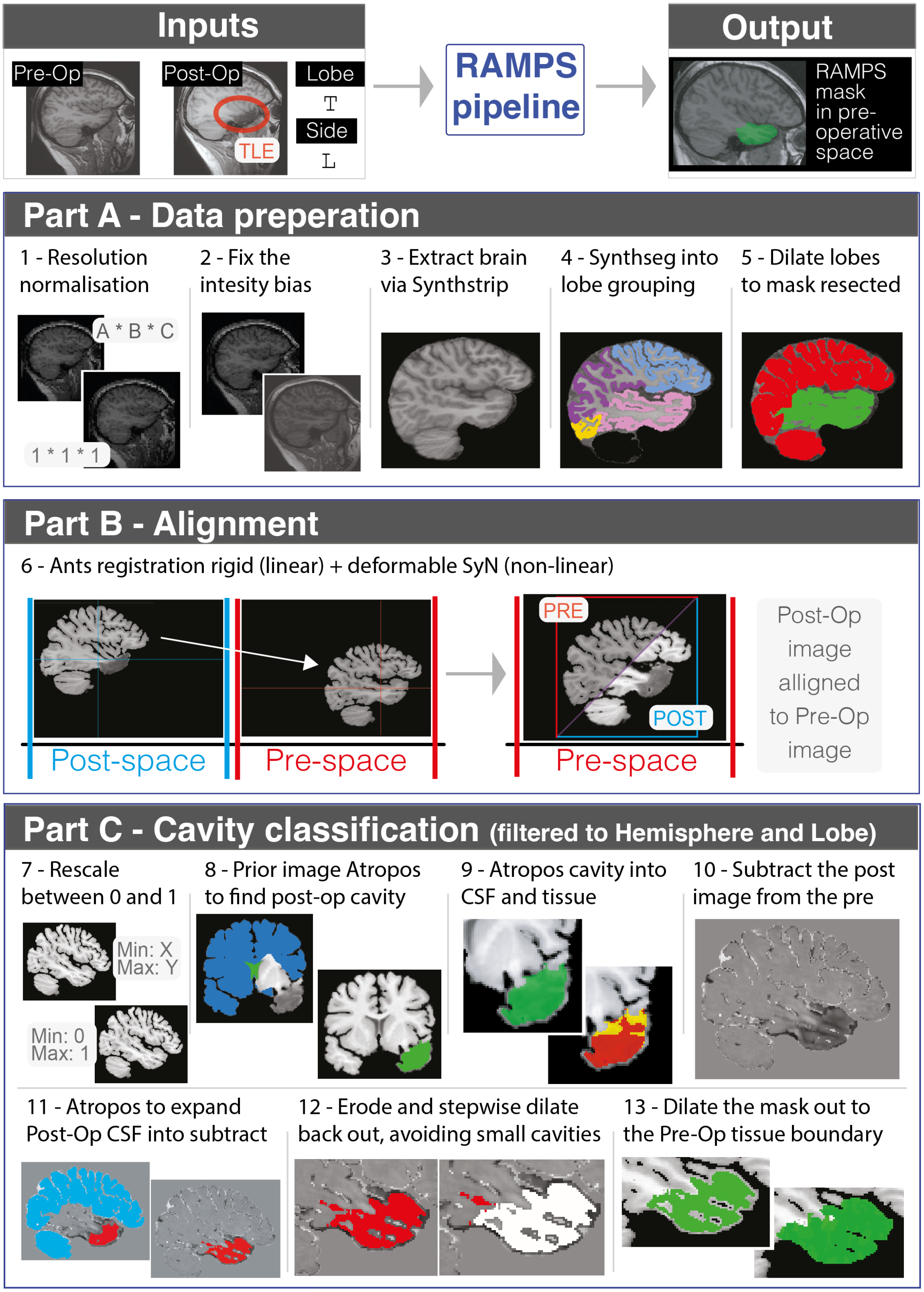

***Figure 2: Overview of the RAMPS pipeline.*** *Using a T1w pre- and post-op image, as well as hemisphere and lobe of resection, RAMPS generates a mask of resected pre-op tissue in 3 steps. Step 1 is data preparation, in which a series of steps are undertaken to remove noise*

*from the image, extract the brain tissue and create a lobe atlas map of the brain. Step 2 uses ANTS registration to align the post-op brain to the pre-op. Step 3 is mask creation which involves delineating the resection cavity in the post-op space resection and then expanding to the pre-op tissue boundary.*

### 2.3.1 Required parameters

RAMPS requires the following :

- Preoperative image - A T1w image captured before surgery.
- Postoperative image - A T1w image captured after surgery, with a clear surgery.
- Output folder - The folder where the outputs of RAMPS will be stored.

While the code can be run with just this information we recommend providing :

- A subject ID.
- Hemisphere initials where resection took place (L,R or both). Default is both.
- Lobe of resection initial(s), if the resection spans multiple lobes then multiple can be specified. (T, F, O, P). Temporal lobe filter also includes Sub-cortical and insula regions. Default is all lobes.

Hemisphere and lobe inputs to ensure that a mask is generated in the correct area and is assumed as prior knowledge from the end-user.

### 2.3.2 Part A - Data preparation

To enhance image alignment and improve the ease of creating a resection mask there are a series of data preparation steps to remove noise and generate lobe classifications masks.

- **Step 1 - Resolution normalisation.** First apply the ANTS (Tustison et al., 2021) resample_image_to_target function to resample images into 1 x 1 x 1 mm resolution with 265 * 256 * 256 fov. This keeps the images in their original space while standardizing FOV and voxel size.

- **Step 2 - Bias intensity correction.** Images are then ran through N4BiasFieldCorrection (Tustison et al., 2010) to correct any low frequency intensity non-uniformity present in MRI image data known as a bias or gain field. This corrects the slow and smooth intensity variation across the image, thereby reducing field bias. At the end this section, to further ensure bias removal, the top 1% voxels values were replaced by the median voxel value.

- **Step 3 - Brain extraction.** Removal of non-brain elements (i.e skull, eyes etc) reduces potential complications in the registration step, especially around the resection cavity (Casseb et al., 2021). The skull-stripping tool SynthStrip is used remove non-brain elements from the T1w images. SynthStrip has been shown to be a robust model and agnostic to acquisition specifics (Hoopes et al., 2022; Kelley et al., 2024).

- **Step 4 - Regional segmentation.** Through the use of SynthSeg (Billot et al., 2023), the images are segmentated into a series of atlas regions. SynthSeg is robust across various brain scans of differing contrast and resolution. Atlas regions are then joined via lobe to create a grey-matter lobe atlas map of the following regional categories: Frontal, Parietal, Temporal, Occipital, Insula, Sub-Cortical areas and areas in which the resection cannot occur (such as ventricles, brainstem and cerebellum). Additionally, the SynthStrip brain image is multiplied by a binarised mask made from the SynthSeg segmented atlas, The rationale is to eliminate any remaining non-brain elements or residual surface left in the image, particularly around the resection cavity

- **Step 5 - Lobe of resected area.** The grey-matter lobe atlas map is dilated throughout the white matter to create a full lobe map. The dilated atlas is subsequently divided into two binary masks based on users specification: the hemisphere-specific lobe where the resection occurred, and the other lobes where the resection did not occur.

### 2.3.3 Part B - Post-op image alignment into pre-op image space registration

- **Step 6 - Registration.** At this stage, the T1w images have been cleaned but still exist in different coordinate spaces. Before creating the resection mask, align the tissue in the post-op image to the pre-operative image. Due to the sagging and swelling that may be seen in post-operative image, this alignment is critical, as poor alignment could erroneously cause further misalignment of tissue. The ANTS antsRegistrationSyN rigid + deformable syn approach, was found to be accurate in the presence of sagging and swelling. In addition the calculated transformations are used to move information created from the post-op image into the pre-op space.

### 2.3.4 Part C - Cavity classification

Following registration, the post-op image exists in the pre-op space and the following steps produce a resection mask in pre-op space.

- **Step 7 - Rescale.** To contrast the pre- and post-op images against one another, first rescale the images between 0 and 1, to ensure that the cerebrospinal fluid (CSF), grey matter and white matter exhibit similar voxel values across images.
- **Step 8 - Post-op image atropos.** Next, delineate the resection cavity within the post-operative image. This is achieved through ANTs Atropos, a tissue segmentation tool ran with k-means clustering (Avants et al., 2011). The clusters are produced by using a prior image, a set of classification masks in which clusters are partially defined. The prior image consists of 1) ventricles for classification of cerebrospinal fluid (CSF) and 2) tissue in the non-resected lobes. Through this method, the lobe is segmented into intact tissue and the resected cavity.
- **Step 9 - Post-op image find cavity CSF.** While the previous step identifies the resection cavity in the post-operative image, it will likely include the resection and also some damaged tissue. Atropos classification is reapplied to further segment the resection cavity into these components.
- **Step 10 - Image subtraction.** Next subtract the post-operative rescaled image from the pre-operative rescaled image to create a difference image. This approach is based on the rationale that after step 7, subtraction of the same voxel class (grey matter, white matter and CSF) will roughly equal zero, whereas overlap of differing classes will be not equal zero. This highlights the overlap between the resection cavity observed in the post-operative image (CSF) and tissue in the pre-operative image, and indicates where sagging has occurred within the image. The resulting subtraction image is then multiplied by the mask of pre-operative image, highlighting the areas of tissue difference in the pre-operative image.
- **Step 11 - Atropos through subtraction image.** Similar to step 8, Atropos is used with a prior image to expand the post-op resection cavity through the subtraction image. This highlights the voxels in the subtracted images where the images differ. Additionally, the results of this Atropos is filtered to the lobe in which resection take place and the largest object is selected to be the mask of the resection cavity.
- **Step 12 - Cavity removal.** A common issue with the subtraction image arises from poor registration, leading to differences caused by tissue misalignment and not resection. These sections of poor alignment are often attached to the main resection mask but via narrow contact points of a few voxels. These areas of noise can be removed by first eroding the mask and performing a series of small dilations through the original mask. After each dilation we examine the voxels expanded into. If expansion reveals a small cluster of voxels, it indicates that there will be an area of poor alignment. Further dilation into this region is prevented, re-creating the step 11 mask, but removing these areas of misalignment.

- **Step 13 - Boundary dilation.** To ensure that the mask extends to the appropriate tissue boundary a directional dilation is preformed. If a given voxel on the boundary of the mask is within 3 voxels of CSF all voxels between the two points onto the mask are added. This stops the dilation into tissue that was not resected.

- **Step 14 - Additional cleaning.** ANTs morphology is applied to fill any small cavities within the resection mask. Then the mask is multiplied by a binarized mask of the pre-op lobe of resection to remove any voxels that might exist outside this region. Finally the resection mask is resampled back into the pre-op resolution.

## 2.4 Data analysis

The primary analysis evaluated the masks produced via RAMPS against the manually drawn cohort. First, the dice similarity coefficient (DSC) between the RAMPS and manual masks across the TLE and ETLE cohort was computed. Additionally the miss rate/false discovery rate of the RAMPS mask produced to highlight how close they are to manually drawn mask was examined. The secondary analysis compared the results of RAMPS to thosefrom additional automation tools (Epic-CHOP and ResectVol) using both DSC and miss rate/false discovery rate approaches to evaluate the effectiveness of RAMPS.

## 2.5 Statistical analysis

As in other studies (Arnold et al., 2022; Casseb et al., 2024; Casseb et al., 2021; Courtney et al., 2024; Wilke et al., 2011), we compared the RAMPS resection mask to manually drawn masks using the Dice Similarity Coefficient (DSC). The DSC has shown to be a robust metric of both overlap and reproducibility, ranging from 0 (no overlap) to 1 (which in this study indicates a 1 to 1 replication of the manually drawn mask) (Fyllingen et al., 2016). DCS between 0 to 0.6 is poor, 0.6 to 0.7 is good, 0.7 to 0.8 is considered high and values exceeding 0.8 are excellent (Seghier et al., 2008). We calculated the median DSC and the interquartile range (IQR).

Additionally, the overlap of the two masks were viewed as a confusion matrix to extract additional metrics for analysis, and to provide a more complete representation of alignment. The voxels which feature in the two masks are true positives, voxels that only exist in the manual are false negatives, voxels that only exist in the automated masks are false positives and voxels that are outside both masks are true negatives. Using these definitions, we can calculate the miss rate, which quantifies the percentage of resected tissue not included by the automated mask and false discovery rate (FDR), a percentage measure of the automated mask that included extra non-resected tissue. The optimal result would be 100% overlap with 0% in Miss rate and FDR, this indicates a perfect overlap between masks.

## 2.6 Comparison to other pipelines

The effectiveness of RAMPS was compared with two other state of the art tools that aim to generate a resection mask in the pre-operative space, Epilepsy Cavity Characterisation Pipeline (Epic-CHOP) (Cahill et al., 2019) and ResectVol (Casseb et al., 2021). Both of these pipelines find the resection cavity using Statistical Parametric Mapping (SPM12) software (Friston et al., 1994) and operate within MATLAB. Both tools follow a similar approach to mask creation, aligning the pre and post-operative images and then finding the differences between the two. While the RAMPS process is similar, the three pipelines use different registration techniques, vary in their operational sequences and employ different methods to generate the final resection mask.

ResectVol and Epic-CHOP were executed using the default values, though as no PET images exists for these subjects Epic-CHOP was filtered to only non-PET steps. By default Epic-CHOP produces a series of possible resection masks. We evaluate the largest cluster mask produced which underwent all cleaning steps. While the possibility exists that another generated masks could more accurately delineate the resection cavity it has been shown Epic-CHOP provides no significant difference between the median DSCs of the automated and semi-automated versions of Epic-CHOP (Courtney et al., 2024). Automated masks were converted to $1mm^3$ resolution and filtered by a binary segmentation of the pre-operative brain tissue to keep the masks within the tissue to match the manual mask.

Epic-CHOP and ResectVol were performed on the same cohort of 87 subjects pre- and post-operative scans; however, some cases could not be processed by the additional pipelines with their default set up, producing hard errors on computation. Thus, we compared the output in the 62 subjects whose data were usable for all three pipelines.

## 2.7 Ethics

The studies involving human participants were reviewed and approved by Newcastle University Research Office Ref: 1804/2020.

## 2.8 Data & code availability

Raw preoperative MRI scans are available as part of the IDEAS dataset (Taylor et al 2024). Code is available at the following location : https://github.com/cnnp-lab/RAMPS

# 3 Results

## 3.1 RAMPS performance using the dice similarity coefficient (DSC)

The results from this primary analysis are presented with mask examples in figure 3a for TLE and figure 3b for ETLE. Comparing RAMPS to manually drawn masks, in the TLE cohort there was an an excellent median DSC of 0.86 (IQR, 0.078) and a high median DSC of 0.71 (IQR, 0.32) in the ETLE cohort. These results are similar to the inter-rater variability in pre-operative mask creation (Sup 8). One mask in the TLE group was poor (DSC $<$ 0.6), as were 6 in the ETLE group. TLE had 68 cases (97%) and ETLE had 10 (53%) cases with high DSC (DSC >= 0.7).

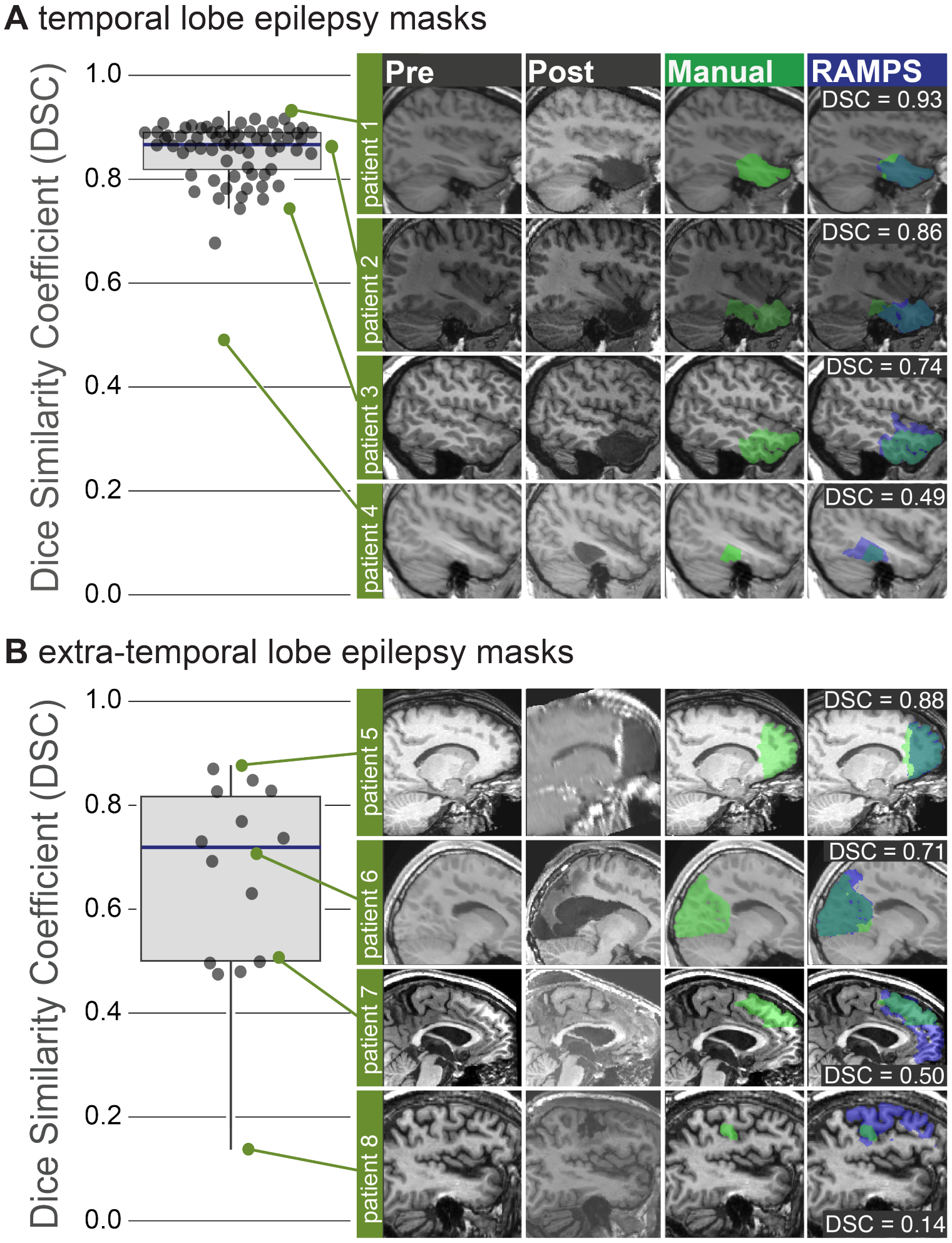

***Figure 3: Similarity of manual and RAMPS generated masks.*** *Using the Dice similarity coefficient for mask comparison, the TLE cohort (A) achieved a median similarity of 0.86 (IQR, 0.078) and the ETLE cohort (B) achieved a median overlap of 0.71 (IQR, 0.32). Example images, along with generated masks are shown in right panels.*

## 3.2 RAMPS evaluation under different performance metrics

Masks generated for TLE had a higher overlap (median 76% TLE, 55% ETLE), lower false discovery rate (median 16% TLE, 40% ETLE), and lower miss rate (median 7% TLE, 7% ETLE) than for ETLE (Fig 4).

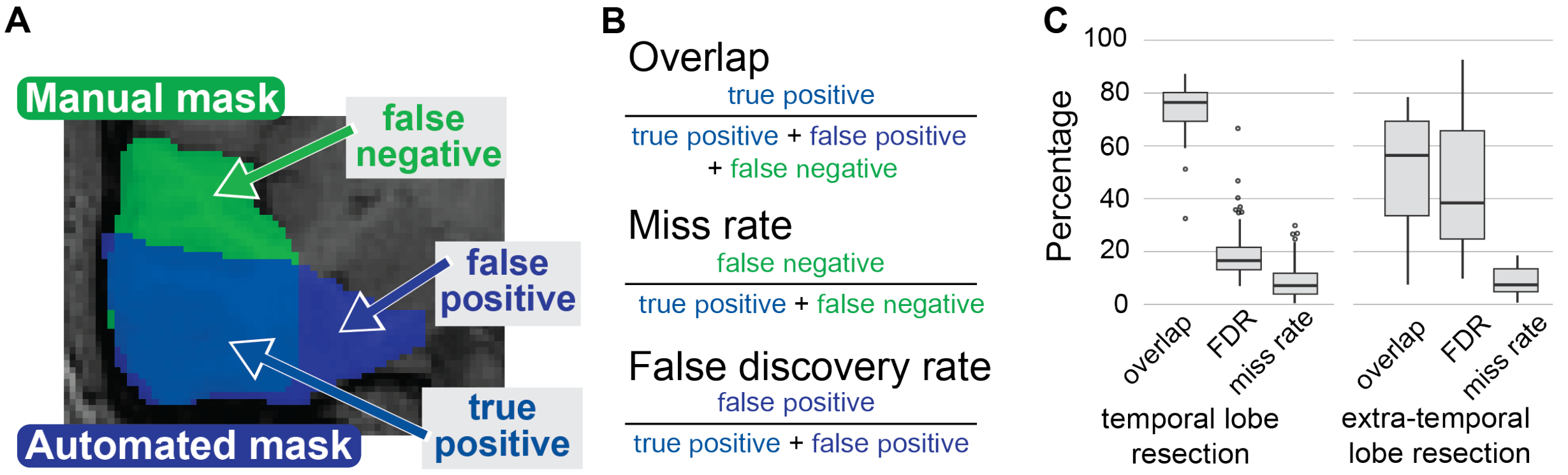

***Figure 4: Performance metrics to analyse mask similarity besides DSC*** *A) Example of a possible manual and an automated mask. This overlap can be divided into three categories: true positive, which are voxels included in both masks; false negatives, which are voxels included only in the manual mask; and false positives, which are voxels included only in the automated mask. B) Using these metrics, the following may be calculated: Overlap, which measures how effective the automation is at capturing the manual mask; Miss rate, representing the percentage of the manual mask not captured; and False discovery rate (FDR), indicating the percentage of voxels identified by automation that fall outside the Manual mask. C) Shows the results of the comparison of RAMPS to manual under these metrics.*

## 3.3 Comparison of different pipelines

It was possible to process 62 across the default set up of both additional pipelines (Figure 5).

RAMPS achieved the highest DSC median(IQR) at 86%(7%), with Epic-CHOP and ResectVol achieving 72%(16% and 21% respectively) (figure 5a). For median(IQR) overlap, RAMPS had the highest at 76%(11%), Epic-CHOP achieved 56%(21%) and ResectVol achieved 56%(24%)(figure 5b). The optimal miss rate is zero and RAMPS achieves the best median at 7%(9%) with both Epic-CHOP and ResectVol scoring 38%(21% and 25% respectively). The optimal False discovery rate is 0, Epic-CHOP was best with a median(IQR) FDR at 4%(12%), ResectVol 7%(8%) and RAMPS at 17%(11%). The inference from the Miss rate and FDR is that the RAMPS masks produced by RAMPS are generally larger than the manually drawn mask, whereas Epic-CHOP and ResectVol are typically smaller and hence have fewer false discoveries.

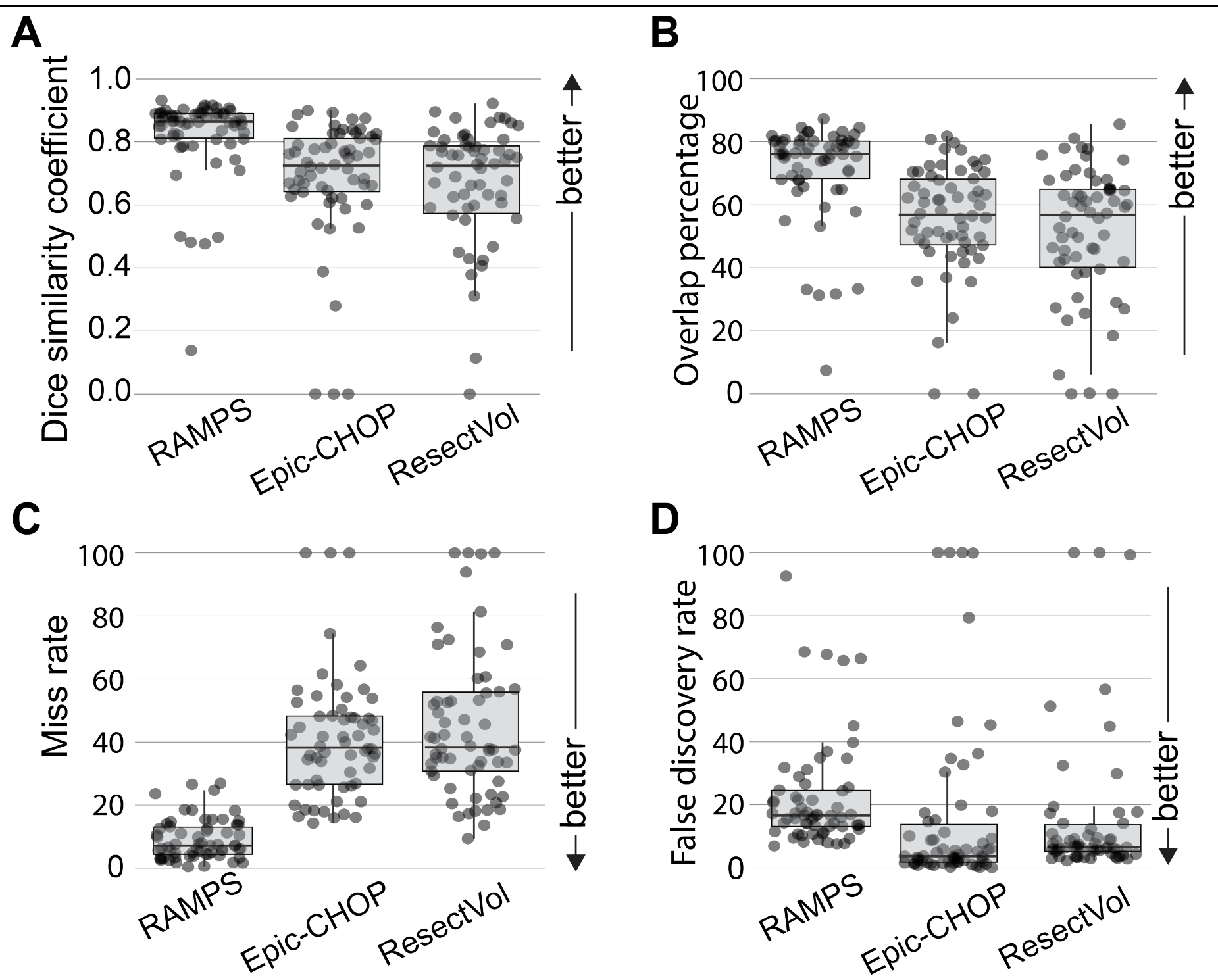

***Figure 5: Pipeline comparison.*** *The automated masks produced by RAMPS, Epic-CHOP and ResectVol compared against a cohort of manual masks (N=62). Each data point represents an individual patient. The median(IQR) were: A) DSC metrics: RAMPS 86%(7%) Epic-CHOP 72%(16%) ResectVol 72%(21%) B) Overlap metrics: RAMPS 76%(11%) Epic-CHOP 56%(21%) ResectVol 56%(24%) C) Miss rate: RAMPS 7%(9%) Epic-CHOP 38%(21%) ResectVol 38%(25%) D) False discovery rate: RAMPS 17%(12%) Epic-CHOP 4%(12%) ResectVol 7%(8%)*

## 3.4 RAMPS use with unusual cases

Finally, we tested RAMPS in cases that differed from the main cohort, including a small lesionectomy, a second surgery, preoperative cavernoma and extreme sagging into the CSF cavity (Figure 6). Although quantitative metrics were not calculated for these isolated cases due to the lack of a manual resection mask, the visual results further demonstrate RAMPS' ability to perform segmentation across a diverse range of cases.

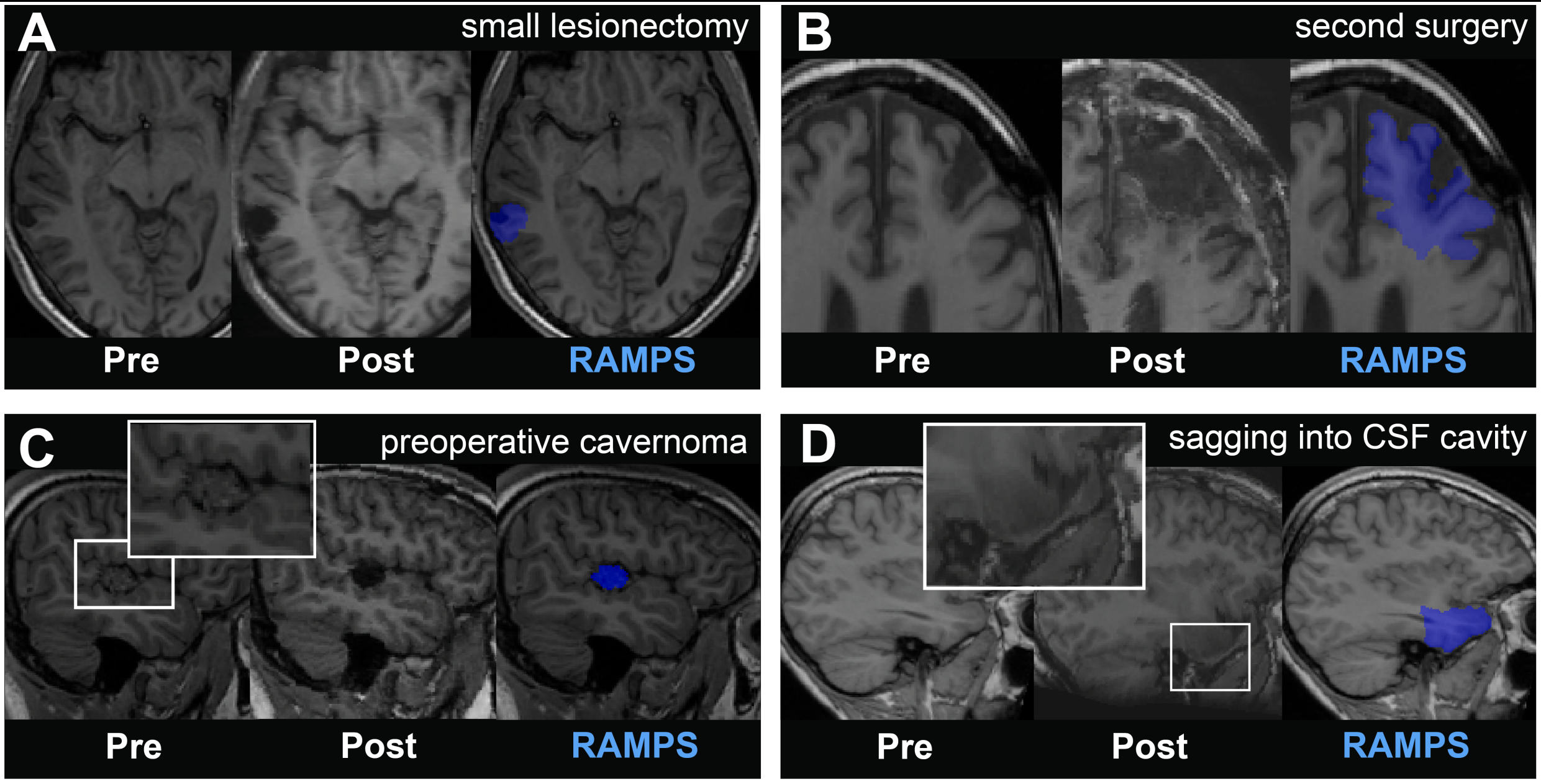

***Figure 6: Demonstrating RAMPS' ability to generate masks under nonstandard situations*** *Here we visualize the automated masks produced for A) a small lesionectomy B) a second surgery C) preoperative cavernoma and D) extreme sagging into the CSF cavity.*

## 4 Discussion

Accurate delineation of resected tissue is of high importance in the study of epilepsy surgery. Most studies delineate tissue manually; however, this is costly in time and expertise and varies between raters. We present RAMPS, a fully automated tool for delineation of the pre-operative tissue that was found to be resected within the post-op image. We compared RAMPS automated masks against 87 manually drawn masks gathered from a diverse cohort. We obtained excellent mask similarity with manual masks in TLE (median DSC=0.86) and a high DSC in ETLE (median DSC=0.71). As inter-rater variability is DSC 0.8 (Sup 8), these results suggest RAMPS is able to segment to the level of a human rater.

The dice similarity coefficient is widely used in this type of analysis (Arnold et al., 2022; Casseb et al., 2024; Casseb et al., 2021; Courtney et al., 2024; Wilke et al., 2011). However, when DSC is less than 1 it does not inform *how* the masks are misaligned. To overcome this limitation, we examined the percentage differences between the automated masks, in terms of overlap, miss rate and false discovery rate. In the TLE cohort RAMPS had a median(IQR) overlap of 76%(12%), a miss rate of 7%(8%) and a FDR of 16%(9). This highlights that RAMPS was able to capture 93% of the manual mask with misalignment primarily coming from over extension in some areas. In the ETLE cases the Overlap median was only 54%(37%); but, RAMPS still captured 90% of the manual masks volume. The use of additional metrics gave a more complete picture of the algorithm's success.

RAMPS transformed the time expensive task of resection cavity delineation to an automated computational process. This is beneficial when needing to analyse a cohort of images, and the mask creation can be sped up through additional computational resources. RAMPS is capable of generating masks on par with a manual rater and, as seen in Figure 3, RAMPS may classify the pre-operative tissue better than the manual rater in a few cases (Figure 3 patient 4,6,8). Users, however, should still visually review the masks for accuracy. As indicated by the miss rate seen in figure 4, a majority of the gold standard mask exists within the RAMPS mask so it may be quickly trimmed by human raters to match the manual evaluation of the resection.

Other studies have shown effectiveness of automatically generating a resection mask (Arnold et al., 2022; Casseb et al., 2024; Courtney et al., 2024; Wilke et al., 2011). However, only few create the mask in pre-operative space. This generation space is crucial as most studies will analyse pre-operative data. Here we compared two previously established pipelines with RAMPS: Epic-CHOP (Cahill et al., 2019) and ResectVol (Casseb et al., 2021). RAMPS achieved both a higher median DSC and median overlap than both ResectVol and Epic-CHOP. Both these pipelines had a miss rate of 38%, which is higher than RAMPS FDR and miss rate combined, meaning more effort would be needed to bring these automated masks to the gold standard. For this analysis all automated masks were filtered within a brain segmentation mask to examine delineation solely within the brain. Without this step the other pipelines masks could include additional non-cerebal voxels, further increasing the false discovery rate and lowering DSC. Additionally, for some subjects these pipelines did not successfully produce a mask in the correct location (i.e. DSC = 0).

Translation of the post-op cavity into pre-op space requires finding the differences between the two images. Therefore importance should primarily be placed on image alignment. Poor translation may not re-align any sagging or swelling within tissue around the resection cavity to its pre-operative position, effecting the ability to accurately segment the resection cavity.

A limitation of this study is that we compared the automatically generated masks to only one manually drawn mask. Since there is variability between manual raters, there is no clear gold standard for comparison. Conversely, this highlights a potential advantage of automated approaches as it is possible that those generated automatically are more accurate than the manual masks used for comparison. We would always recommend visual inspection of any mask created.

Another limitation of RAMPS is the optional request for the user to input additional information regarding the hemisphere and lobe of resection for optimal usage, as other pipelines can operate without this request. As this information can be easily extracted through visual inspection, or from clinical notes this is not expected to be a burden on users. This inclusion ensures RAMPS creates a mask in the correct tissue area, as shown with RAMPS capturing the cavity (DSC $> 0$) in all cases. The RAMPS pipeline, however can be used without such information, though the outputs would be expected to be less accurate.

While we show RAMPS's effectiveness in a diverse cohort, development depended on the data we had available. In future we would aim to extend the pipeline for (i) Laser Interstitial Thermal Therapy (LiTT) ablations (ii) disconnections (iii) alternative imaging besides T1w MRI, specifically post-operative CT. We release RAMPS as an open source tool and encourage users to modify RAMPS to meet individuals situations and needs.

In summary, RAMPS is a new pipeline to generate masks of a resected cerebral cavity seen in the post-operative image, registered to the tissue resected in the pre-operative image. We provide RAMPS as a free, open source, robust, easy to use tool, which we anticipate will be particularly useful for large datasets (Taylor et al., 2024).

## Author contributions

Conceptualisation: CS, GH, YW, PT. Methodology: CS, GH, YW, PT. Software: CS, GH, PT. Validation: CS, GH, YW, PT. Investigation: CS, GH, YW, PT. Resources: JSD, YW, PT. Data curation CS, GH, JSD, YW, PT. Writing – Original draft: CS. Writing – review and editing: CS, GH, JSD, YW, PT. Visualisation: CS, PT. Supervision: PT. Funding acquisition: YW, JSD, PT

## Acknowledgements

We thank members of the Computational Neurology, Neuroscience & Psychiatry Lab (www.cnnp-lab.com) for discussions on the analysis and manuscript; P.N.T. and Y.W. are both supported by UKRI Future Leaders Fellowships (MR/T04294X/1, MR/V026569/1). JSD, JdT are supported by the NIHR UCLH/UCL Biomedical Research Centre.

## Competing interests

The authors declare no competing interests.

## Supplementary

## Interrater variability

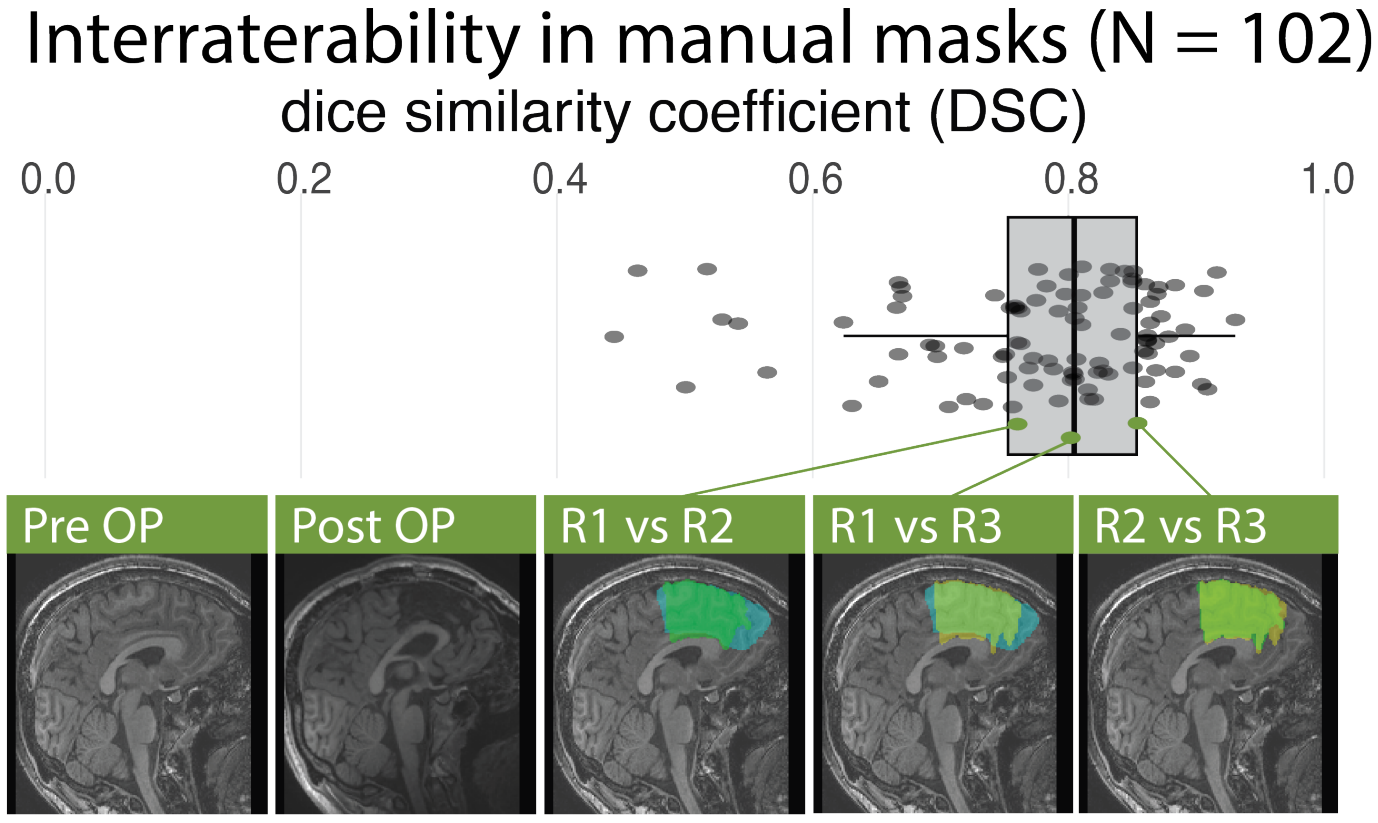

***Figure S1: Intra-rater variability in resection mask creation.*** *To examine if differences between manual raters exists, the dice score similarity (DSC) was calculated comparing the resection mask from two different raters classifying the same subject. In total 103 comparisons where carried out to discover the median(IQR) DSC between two raters is 0.8(0.1).*

## Direct comparison of pipelines

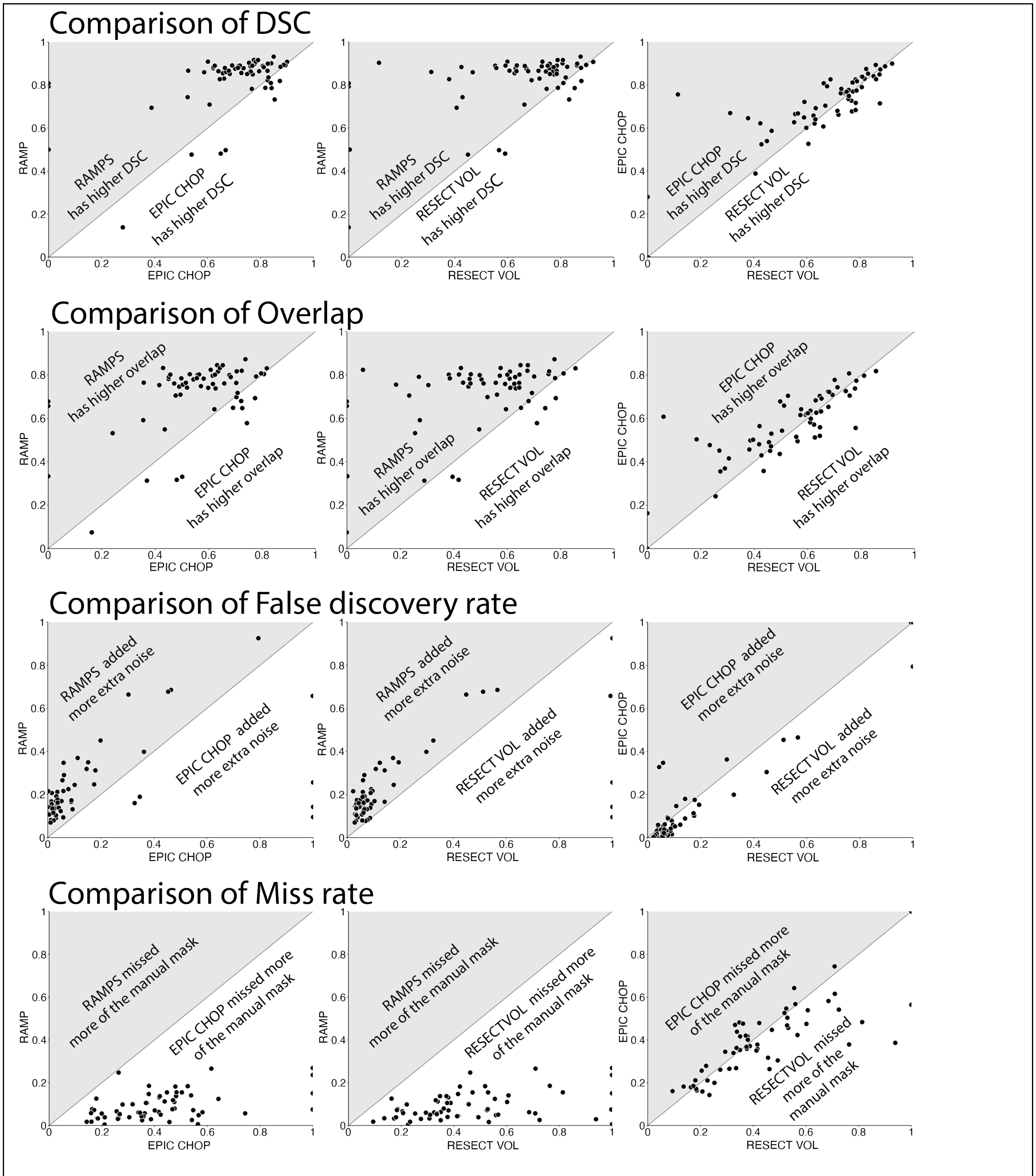

***Figure S2 Comparison between pipelines.*** *Examination of the pipelines performance against each another under DSC, Overlap, false discovery rate and miss rate. Each data point is a mask.*

## References

Ahmed, B., Brodley, C.E., Blackmon, K.E., Kuzniecky, R., Barash, G., Carlson, C., Quinn, B.T., Doyle, W., French, J., Devinsky, O., others, 2015. Cortical feature analysis and machine learning improves detection of "MRI-negative" focal cortical dysplasia. Epilepsy & Behavior 48, 21–28.

Arnold, T.C., Muthukrishnan, R., Pattnaik, A.R., Sinha, N., Gibson, A., Gonzalez, H., Das, S.R., Litt, B., Englot, D.J., Morgan, V.L., others, 2022. Deep learning-based automated segmentation of resection cavities on postsurgical epilepsy MRI. NeuroImage: Clinical 36, 103154.

Avants, B.B., Tustison, N.J., Wu, J., Cook, P.A., Gee, J.C., 2011. An open source multivariate framework for n-tissue segmentation with evaluation on public data. Neuroinformatics 9, 381–400.

Billardello, R., Ntolkeras, G., Chericoni, A., Madsen, J.R., Papadelis, C., Pearl, P.L., Grant, P.E., Taffoni, F., Tamilia, E., 2022. Novel user-friendly application for MRI segmentation of brain resection following epilepsy surgery. Diagnostics 12, 1017.

Billot, B., Greve, D.N., Puonti, O., Thielscher, A., Van Leemput, K., Fischl, B., Dalca, A.V., Iglesias, J.E., others, 2023. SynthSeg: Segmentation of brain MRI scans of any contrast and resolution without retraining. Medical image analysis 86, 102789.

Brett, M., Markiewicz, C.J., Hanke, M., Côté, M.-A., Cipollini, B., Papadopoulos Orfanos, D., McCarthy, P., Jarecka, D., Cheng, C.P., Larson, E., Halchenko, Y.O., Cottaar, M., Ghosh, S., Wassermann, D., Gerhard, S., Lee, G.R., Baratz, Z., Moloney, B., Wang, H.-T., Kastman, E., Kaczmarzyk, J., Guidotti, R., Daniel, J., Duek, O., Rokem, A., Scheltienne, M., Madison, C., Sólon, A., Morency, F.C., Goncalves, M., Markello, R., Riddell, C., Burns, C., Millman, J., Gramfort, A., Leppäkangas, J., Bosch, J.J.F. van den, Vincent, R.D., Braun, H., Subramaniam, K., Van, A., Legarreta, J.H., Gorgolewski, K.J., Raamana, P.R., Klug, J., Vos de Wael, R., Nichols, B.N., Baker, E.M., Koudoro, S., Hayashi, S., Pinsard, B., Haselgrove, C., Hymers, M., Esteban, O., Pérez-García, F., Becq, G., Dockès, J., Oosterhof, N.N., Amirbekian, B., Christian, H., Nimmo-Smith, I., Nguyen, L., Suter, P., Reddigari, S., St-Jean, S., Panfilov, E., Garyfallidis, E., Varoquaux, G., Newton, J., Hahn, K.S., Waller, L., Hinds, O.P., Sandro, Fauber, B., Dewey, B., Perez, F., Roberts, J., Poline, J.-B., Stutters, J., Jordan, K., Cieslak, M., Moreno, M.E., Hrnčiar, T., Haenel, V., Schwartz, Y., Darwin, B.C., Thirion, B., Gauthier, C., Solovey, I., Gonzalez, I., Palasubramaniam, J., Lecher, J., Leinweber, K., Raktivan, K., Calábková, M., Fischer, P., Gervais, P., Gadde, S., Ballinger, T., Roos, T., Reddam, V.R., freec, 2024. Nipy/nibabel: 5.3.1.

Cahill, V., Sinclair, B., Malpas, C.B., McIntosh, A.M., Chen, Z., Vivash, L.E., O'Shea, M.F., Wilson, S.J., Desmond, P.M., Berlangieri, S.U., others, 2019. Metabolic patterns and seizure outcomes following anterior temporal lobectomy. Annals of neurology 85, 241–250.

Casseb, R.F., Campos, B.M. de, Loos, W.S., Barbosa, M.E.R., Alvim, M.K.M., Paulino, G.C.L., Pucci, F., Worrell, S., Souza, R.M. de, Jehi, L., others, 2024. Fully automatic segmentation of brain lacunas resulting from resective surgery using a 3D deep learning model. medRxiv 2023–11.

Casseb, R.F., Campos, B.M. de, Morita-Sherman, M., Morsi, A., Kondylis, E., Bingaman, W.E., Jones, S.E., Jehi, L., Cendes, F., 2021. ResectVol: A tool to automatically segment and characterize lacunas in brain images. Epilepsia Open 6, 720–726.

Courtney, M.R., Sinclair, B., Neal, A., Nicolo, J.-P., Kwan, P., Law, M., O'Brien, T.J., Vivash, L., 2024. Automated segmentation of epilepsy surgical resection cavities: Comparison of four methods to manual segmentation. NeuroImage 120682.

De Tisi, J., Bell, G.S., Peacock, J.L., McEvoy, A.W., Harkness, W.F., Sander, J.W., Duncan, J.S., 2011. The long-term outcome of adult epilepsy surgery, patterns of seizure remission, and relapse: A cohort study. The Lancet 378, 1388–1395.

Friston, K.J., Holmes, A.P., Worsley, K.J., Poline, J.-P., Frith, C.D., Frackowiak, R.S., 1994. Statistical parametric maps in functional imaging: A general linear approach. Human brain mapping 2, 189–210.

Fyllingen, E.H., Stensjøen, A.L., Berntsen, E.M., Solheim, O., Reinertsen, I., 2016. Glioblastoma segmentation: Comparison of three different software packages. PLoS One 11, e0164891.

Galovic, M., Baudracco, I., Wright-Goff, E., Pillajo, G., Nachev, P., Wandschneider, B., Woermann, F., Thompson, P., Baxendale, S., McEvoy, A.W., others, 2019. Association of piriform cortex resection with surgical outcomes in patients with temporal lobe epilepsy. JAMA neurology 76, 690–700.

Gau, K., Schmidt, C.S., Urbach, H., Zentner, J., Schulze-Bonhage, A., Kaller, C.P., Foit, N.A., 2020. Accuracy and practical aspects of semi-and fully automatic segmentation methods for resected brain areas. Neuroradiology 62, 1637–1648.

Hall, G.R., Hutchings, F., Horsley, J., Simpson, C.M., Wang, Y., Tisi, J. de, Miserocchi, A., McEvoy, A.W., Vos, S.B., Winston, G.P., others, 2023. Epileptogenic networks in extra temporal lobe epilepsy. Network Neuroscience 7, 1351–1362.

Hoopes, A., Mora, J.S., Dalca, A.V., Fischl, B., Hoffmann, M., 2022. SynthStrip: Skull-stripping for any brain image. NeuroImage 260, 119474. https://doi.org/https://doi.org/10.1016/j.neuroimage.2022.119474

Horsley, J.J., Thomas, R.H., Chowdhury, F.A., Diehl, B., McEvoy, A.W., Miserocchi, A., Tisi, J. de, Vos, S.B., Walker, M.C., Winston, G.P., others, 2023. Complementary structural and functional abnormalities to localise epileptogenic tissue. EBioMedicine 97.

Keller, S.S., Richardson, M.P., Schoene-Bake, J.-C., O'muircheartaigh, J., Elkommos, S., Kreilkamp, B., Goh, Y.Y., Marson, A.G., Elger, C., Weber, B., 2015. Thalamotemporal alteration and postoperative seizures in temporal lobe epilepsy. Annals of neurology 77, 760–774.

Kelley, W., Ngo, N., Dalca, A.V., Fischl, B., Zöllei, L., Hoffmann, M., 2024. Boosting skull-stripping performance for pediatric brain images. arXiv preprint arXiv:2402.16634.

Kocsis, Z., Jenison, R.L., Taylor, P.N., Calmus, R.M., McMurray, B., Rhone, A.E., Sarrett, M.E., Deifelt Streese, C., Kikuchi, Y., Gander, P.E., others, 2023. Immediate neural impact and incomplete compensation after semantic hub disconnection. Nature communications 14, 6264.

Pérez-García, F., Dorent, R., Rizzi, M., Cardinale, F., Frazzini, V., Navarro, V., Essert, C., Ollivier, I., Vercauteren, T., Sparks, R., others, 2021. A self-supervised learning strategy for postoperative brain cavity segmentation simulating resections. International Journal of Computer Assisted Radiology and Surgery 16, 1653–1661.

Pérez-García, F., Rodionov, R., Alim-Marvasti, A., Sparks, R., Duncan, J.S., Ourselin, S., 2020. Simulation of brain resection for cavity segmentation using self-supervised and semi-supervised learning, in: Medical Image Computing and Computer Assisted Intervention–MICCAI 2020: 23rd International Conference, Lima, Peru, October 4–8, 2020, Proceedings, Part III 23. Springer, pp. 115–125.

Sainburg, L.E., Morgan, V.L., 2024. Investigation of network reorganization after epilepsy surgery is worth the effort. Brain 147, 2261–2263.

Seghier, M.L., Ramlackhansingh, A., Crinion, J., Leff, A.P., Price, C.J., 2008. Lesion identification using unified segmentation-normalisation models and fuzzy clustering. Neuroimage 41, 1253–1266.

Taylor, P.N., Sinha, N., Wang, Y., Vos, S.B., Tisi, J. de, Miserocchi, A., McEvoy, A.W., Winston, G.P., Duncan, J.S., 2018. The impact of epilepsy surgery on the structural connectome and its relation to outcome. NeuroImage: Clinical 18, 202–214.

Taylor, P.N., Wang, Y., Simpson, C., Janiukstyte, V., Horsley, J., Leiberg, K., Little, B., Clifford, H., Adler, S., Vos, S.B., others, 2024. The imaging database for epilepsy and surgery (IDEAS). Epilepsia.

Tustison, N.J., Avants, B.B., Cook, P.A., Zheng, Y., Egan, A., Yushkevich, P.A., Gee, J.C., 2010. N4ITK: Improved N3 bias correction. IEEE transactions on medical imaging 29, 1310–1320.

Tustison, N.J., Cook, P.A., Holbrook, A.J., Johnson, H.J., Muschelli, J., Devenyi, G.A., Duda, J.T., Das, S.R., Cullen, N.C., Gillen, D.L., others, 2021. The ANTsX ecosystem for quantitative biological and medical imaging. Scientific reports 11, 9068.

Wilke, M., Haan, B. de, Juenger, H., Karnath, H.-O., 2011. Manual, semi-automated, and automated delineation of chronic brain lesions: A comparison of methods. NeuroImage 56, 2038–2046.